\documentstyle[aps,preprint,prb]{revtex}
\tighten
\begin{document}
\draft
\newcommand{\ie}{{\em i.e.}}
\newcommand{\eg}{{\em e.g.}}
\newcommand{\vs}{{\em vs.}}
\newcommand{\rme}{\text{e}}
\newcommand{\rmd}{\text{d}}
\preprint{cond-mat/9512110}
\title{
Analytical and computational study of magnetization switching in
kinetic Ising systems with demagnetizing fields
}
\author{Howard L.\ Richards}
\address{
  Center for Materials Research and Technology, \\
  (http://www.martech.fsu.edu) \\
  Department of Physics, \\
  (http://www.physics.fsu.edu) \\
  and Supercomputer Computations Research Institute, \\
  (http://www.scri.fsu.edu)  \\
  Florida State University, Tallahassee, Florida 32306-3016
        }
\address{
  Department of Solid State Physics, \\ Ris{\o}~National Laboratory,
  DK-4000 Roskilde, Denmark \\
  (http://risul1.risoe.dk)
        }
\author{M.~A.\ Novotny}
\address{
  Supercomputer Computations Research Institute,  \\
  Florida State University, Tallahassee, Florida 32306-4052
        }
\address{
  Department of Electrical Engineering,  \\
  (http://eesun3.eng.fsu.edu) \\
  2525 Pottsdamer Street \\
  Florida A\&M University--Florida State University,
  Tallahassee, Florida 32310-6046
        }
\author{Per Arne Rikvold}
\address{
  Center for Materials Research and Technology,
  Department of Physics, \\
  and Supercomputer Computations Research Institute,  \\
  Florida State University, Tallahassee, Florida 32306-3016
         }
\address{
  Centre for the Physics of Materials and Department of Physics, \\
  (http://www.physics.mcgill.ca) \\
  McGill University, Montr{\'e}al, Qu{\'e}bec, Canada
        }
\date{\today}
\maketitle
\begin{abstract}
An important aspect of real ferromagnetic particles is the
demagnetizing field resulting from magnetostatic dipole-dipole
interaction, which causes large particles to
break up into domains.  Sufficiently small particles,
however, remain single-domain in equilibrium.  This makes such
small particles of particular interest as materials for
high-density magnetic recording media.
In this paper we use analytic arguments and
Monte Carlo simulations to study the effect of the
demagnetizing field on the dynamics of magnetization
switching in two-dimensional, single-domain, kinetic Ising systems.
For systems in the ``Stochastic Region,'' where
magnetization switching is on average
effected by the nucleation and
growth of fewer than two well-defined critical droplets,
the simulation
results can be explained by the dynamics of a simple model
in which the free energy is a function only of magnetization.
In the ``Multi-Droplet Region,'' a generalization of Avrami's
Law involving a magnetization-dependent effective magnetic field
gives good agreement with our simulations.
\vspace{1cm}
\begin{center}
	{\bf FSU-SCRI-95-114}
\end  {center}
\end{abstract}

\section{Introduction}
\typeout{Introduction}
\label{sec-intro}

The ability of single-domain ferromagnets to preserve an accurate
record of past magnetic fields has several important applications.
Fine grains in lava flows preserve a record of the direction of
the geomagnetic field at the time they cooled, giving valuable
insight into continental drift and the dynamics of the earth's
core.\cite{DHTarling}
Of more direct technological importance is the potential
application
of single-domain ferromagnets to magnetic recording media,
such as magnetic tapes and disks.

During the magnetic recording process,
different regions of the medium
are briefly exposed to strong magnetic fields, so that each grain
is magnetized in the desired direction.\cite{Koester}
Since each grain can in principle store one bit of data,
a greater storage density could ideally be achieved
by a medium containing many small grains than by
one containing a few large grains.
However, in order to serve as reliable storage devices,
the grains must be capable of retaining their magnetizations
for long periods of time in weaker, arbitrarily oriented
ambient magnetic fields --- \ie, they must have
a high coercivity and a large remanence.
Since experiments show the existence of a
particle size at which the coercivity is maximum
(see, \eg, Ref.~\onlinecite{Kneller63}),
there is a tradeoff between high storage
capacity and long-term data integrity which must give rise
to an optimum choice of grain size for any given material.
During both recording and storage, the relationships between
the magnetic field, the size of the grain, and the lifetime of the
magnetization opposed to the applied magnetic field
are therefore of great technological interest.

Fine ferromagnetic grains have been studied for many
years, but until recently such particles could be
studied experimentally only in powders
(see, \eg, Ref.~\onlinecite{Kneller63}).
This made it difficult to differentiate the statistical
properties of single-grain switching from effects resulting from
distributions in particle sizes, compositions, and local
environments, or from interactions between grains.
Techniques such as magnetic force microscopy (MFM) (see, \eg,
Refs.~\onlinecite{Martin87,Chang93,Lederman93,Lederman94,Lederm94PRL})
and Lorentz microscopy
(see, \eg, Ref.~\onlinecite{Salling93}) now provide
means for overcoming the difficulties in resolving the magnetic
properties of individual single-domain particles.

The standard theory of magnetization reversal
in single-domain ferromagnets is due to N{\'e}el\cite{Neel49}
and Brown.\cite{Brown}
In order to avoid an energy barrier due to exchange interactions
between atomic moments with unlike orientations,
N{\'e}el-Brown theory assumes uniform rotation of
all the atomic moments in the system.
The remaining barrier is caused by magnetic anisotropy,\cite{Jacobs}
which may be either intrinsic or shape-induced.
Anisotropy makes it energetically favorable for each
atomic moment to be aligned along one or more ``easy'' axes.
Buckling, fanning, and curling are, like uniform rotation,
theoretical relaxation processes with few degrees of freedom
and global dynamics.\cite{Koester,Kneller}

\begin{figure}
\vspace*{2.7in}
	 \includegraphics{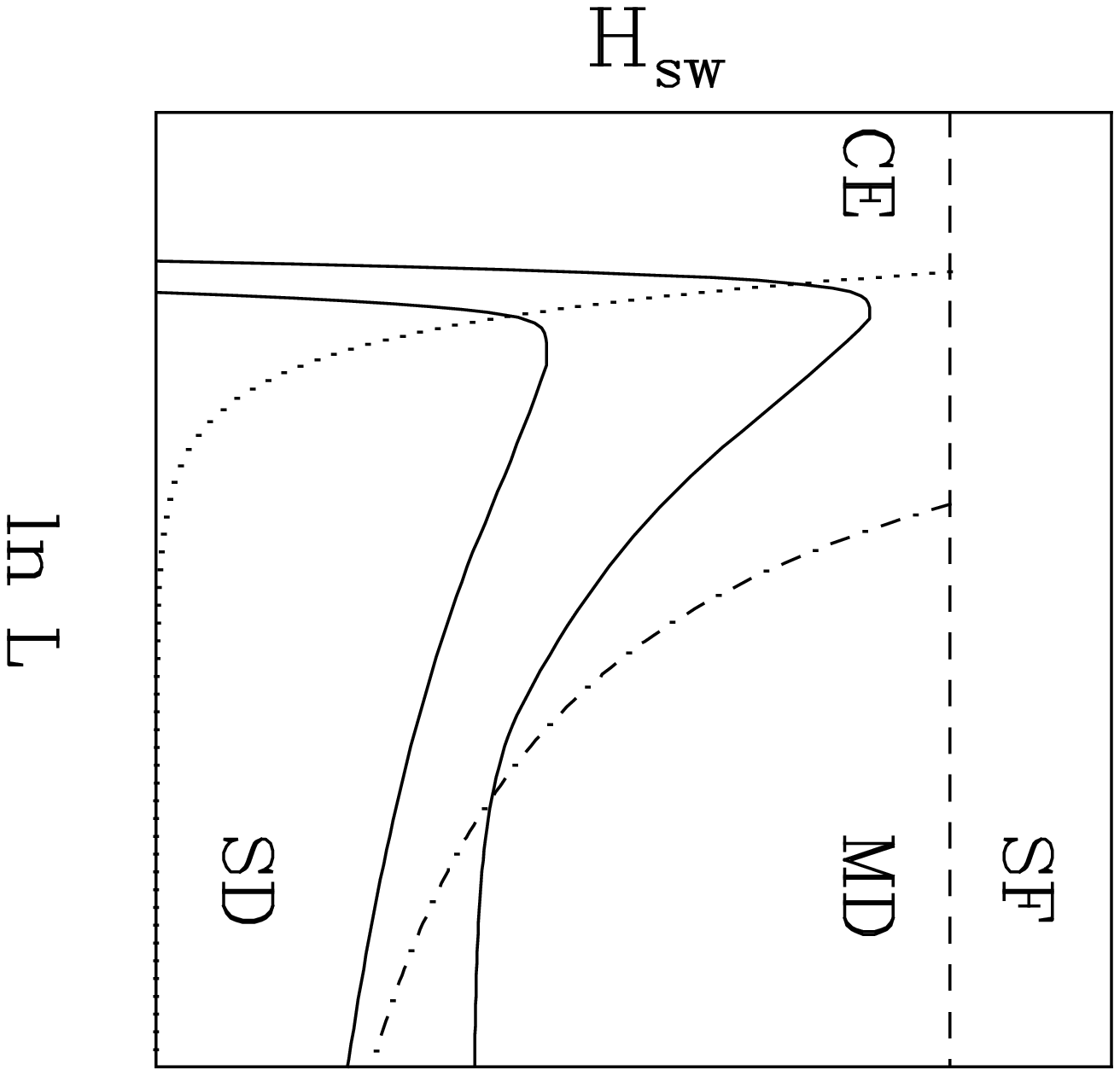}
	  \noindent
	\caption{
    	 	 \label{fig:HsRoad}
		 The relationship between the applied field $H$
		 and system width $L$ for a shorter (top solid curve)
		 and a longer (bottom solid curve) fixed lifetime
		 in a typical metastable
		 magnetic system.
		 Four regions are distinguished by
		 differing decay processes:
		    the Coexistence    region (CE),
		    the Single-Droplet region (SD),
		    the Multi-Droplet  region (MD), and
		    the Strong-Field   region (SF).
		 The CE and SD regions are separated by the
		 thermodynamic spinodal (dotted curve).
		 The SD and MD regions are separated by the
		 dynamic spinodal (dash-dotted curve).
		 The SF region is separated from the other
		 regions by the mean-field spinodal (dashed curve).
			[Reproduced from Fig.~1 of
			Ref.~\protect\onlinecite{2dpi}.]
		}
\end  {figure}

However, for highly anisotropic materials there exists an
alternative mode of relaxation with a typically much shorter lifetime.
Small regions of the phase in which the magnetization is parallel
to the applied magnetic field (the ``stable'' phase)
are continually created and destroyed by thermal fluctuations
within the phase in which the magnetization is antiparallel
to the field (the ``metastable'' phase).
As long as such a region (henceforth referred to as a ``droplet'')
is sufficiently small, the short-ranged exchange interaction
with the surrounding metastable phase imposes a net
free-energy penalty, and the droplet will, with high probability,
shrink and vanish.
Should the droplet become larger than a critical size, however,
this penalty will be less than the benefit obtained from orienting
parallel to the magnetic field, and the droplet will with a
high probability grow further, eventually consuming the grain.
The nature of the metastable decay thus depends on the relative
sizes of the grain, the critical droplet, the average distance
between droplets, and the lattice constant,
as discussed in detail, \eg, in
Refs.~\onlinecite{Orihara92,Tomi92A,Rik94,RikARCP94}.
For systems in which short-range interactions dominate,
Fig.~\ref{fig:HsRoad} sketches the four regions of
the space of magnetic fields and particle sizes 
distinguished by different behaviors during metastable decay.
For a more complete, recent review of droplet theory,
see Ref.~\onlinecite{RikARCP94}.

It is important to understand
the difference between
a droplet and a domain.\cite{Carey}
Although they are both spatially
contiguous regions of uniform magnetization, a domain is an
equilibrium feature whereas {\em a droplet is a strictly
non-equilibrium entity.}
The domain structure of two-dimensional dipole systems has
been extensively
investigated.\cite{Kittel46,Czech89,Kaplan93,MacIsaac95,Booth95,Ng95}
The magnetostatic dipole-dipole interaction
produces a demagnetizing field, which results in the stabilization
of a domain structure in large ferromagnetic particles at
equilibrium.

The purpose of the present paper is to study the effects of
long-range dipole-dipole interactions on the {\em nonequilibrium}
phenomenon of magnetization switching in single-domain
ferromagnetic particles. Towards this end we employ
a simplified model in which particles in equilibrium can
have only one or two domains, and we emphasize the
single-domain case.

Detailed descriptions of both the static and dynamic properties
of fine ferromagnetic grains have typically been formulated
from micromagnetic studies.\cite{WFBrown}
This method involves coarse-graining the physical lattice
onto a computational lattice and then
solving the partial differential equations for the evolution
of magnetic structures on the computational lattice.
Although micromagnetics provides a good treatment for the
anisotropy and demagnetizing fields, it treats thermal effects
rather crudely, usually just by making the domain-wall energy
temperature-dependent.
A somewhat better approximation for thermal fluctuations
within the underlying differential equations is to include
small fluctuations using a Langevin noise term.\cite{Lyberat93}
A better treatment for thermal and time-dependent
effects is therefore Monte Carlo
simulation (see, \eg, Refs.~\onlinecite{Kirby94,Nowak95,Chui95}).
Even when the physical
phenomena can be accurately simulated, however, it will be
difficult to understand the results without an adequate theoretical
basis.

Because of its simplicity, the kinetic nearest-neighbor Ising
model has been extensively studied as a prototype for
metastable dynamics (see Ref.~\onlinecite{RikARCP94} and references
cited therein).
In particular, square- and cubic-lattice Ising
systems with periodic boundary conditions have been used to study
grain-size effects in ferroelectric switching.\cite{Duiker90,Beale93}
A related one-dimensional model has been used to
study magnetization reversal in elongated ferromagnetic
particles.\cite{Braun}
In a recent article,\cite{2dpi}  we applied
statistical-mechanical droplet theory and Monte Carlo simulations
of two-dimensional Ising systems to obtain a qualitative
approximation for the dynamical behavior of real single-domain
particles magnetized opposite to an applied field.  In so doing, we
made several simplifying approximations, one of which was the absence
of a demagnetizing field.
In this article we consider the effect of a small
demagnetizing field on Ising systems, and compare our
analytic calculations with simulations of two-dimensional
Ising systems. Specifically, for systems in the Stochastic
region (discussed in Sec.~\ref{sec-stochastic}), the demagnetizing
fields we consider must be sufficiently small that the system
is consists of a single domain in equilibrium, whereas in the
Multi-Droplet region (discussed in Sec.~\ref{sec-multidrop})
it is sufficient to have for demagnetizing field to be much
smaller than the applied field.
Some preliminary results were
presented in Ref.~\onlinecite{MMM95}.

The organization of this paper is as follows.
In Sec.~\ref{sec-model} we define the model and numerical methods
employed in this paper.
In Sec.~\ref{sec-stochastic} we discuss the Stochastic region
in terms of an approximate free-energy functional and give some
numerical results.
In Sec.~\ref{sec-multidrop} we generalize
Avrami's Law,\cite{Kolmogorov37,JohnsMehl39,Avrami}
which describes magnetization switching in the Multi-Droplet
region, to include the
effects of the demagnetizing field, and we compare the analytical
results to numerical simulations.
Section~\ref{sec-discuss} contains conclusions and discussions.

\section{Model and Numerical Methods}
\typeout{Model and Numerical Methods}
\label{sec-model}
The standard Ising model is defined by the Hamiltonian
\begin{equation}
	\label{eq:Ham_0}
		{\cal H}_0 =
		-J \sum_{\langle i,j \rangle} s_i s_j
		- H L^d m \; ,
\end  {equation}
where $s_i \! = \! \pm 1$ is the $z$-component of the
magnetization of the atom (spin) at site $i,$
$J \! > \! 0$ is the ferromagnetic exchange interaction,
and $H$ is the applied magnetic field times the single-spin
magnetic moment.
The sum $\sum_{\langle i,j \rangle}$ runs over all
nearest-neighbor pairs on a square (generally
$d$-dimensional hypercubic)
lattice of side $L$.
In this work we do not study the effects of grain boundaries,
so periodic boundary conditions are imposed.
The dimensionless system magnetization is given by
\begin{equation}
	\label{eq:sysmag}
		m = L^{-d} \sum_{i} s_i \; ,
\end  {equation}
where the sum is over all $L^d$ sites.
The lattice constant is set to unity.

Addition of dipole-dipole interactions gives a total
Hamiltonian (SI units)
\begin{equation}
  \label{eq:Ham_calD}
	{\cal H}_{\cal D} = {\cal H}_0 + \frac{\mu_0 M^2}{4\pi}
		\sum_{i \neq j}
		\frac{s_i s_j}{|\bbox{r}_{ij}|^3}
		\left[ 1 - 3 \left(
		\frac{\bbox{r}_{ij}}{|\bbox{r}_{ij}|}
		 \cdot \hat{\bbox{ z}}\right)^2 \right] \; ,
\end  {equation}
where $M$ is the saturation magnetic dipole moment density
and $\bbox{r}_{ij}$ is the vector from site $i$ to site $j$.
Unfortunately, however, the last sum in Eq.~(\ref{eq:Ham_calD})
slows down Monte Carlo simulations significantly, which is
problematic if a large number of realizations are desired for
good statistics, as is the case in nonequilibrium
studies.  The last sum also would make a perturbative
expansion in the demagnetizing field (adjustable by changing $M$
or the sample shape) difficult.  We therefore instead use the
simpler Hamiltonian
\begin{eqnarray}
  \label{eq:Ham_D}
	{\cal H}_{D} = {\cal H}_0 + L^d D m^2 \; ,
\end  {eqnarray}
where $D$ is a function of the crystal symmetry, the
shape of the system, and $M$.
Equations~(\ref{eq:Ham_calD}) and (\ref{eq:Ham_D}) are equivalent
for general ellipsoids {\em uniformly}
magnetized along a principal axis.
For the special case of a perpendicularly magnetized plane
with square-lattice symmetry,
$D \! = \! \frac{2}{3} \mu_0 M^2$. For non-uniformly
magnetized systems,
Eq.~(\ref{eq:Ham_D}) amounts to a mean-field treatment of the
effects of the dipole-dipole interactions.

For systems with periodic boundary conditions, the exchange and
dipole terms of Eq.~(\ref{eq:Ham_D}) are
equal when the system size is given by\cite{Kittel46}
\begin{equation}
  \label{eq:L_D}
	L_D \approx \frac{2\sigma_\infty(T)}
                     {D\left[m_{\text{sp}}(T)\right]^2} \; ,
\end  {equation}
where $\sigma_\infty (T)$ is the surface tension along a
primitive lattice vector in the limit $L \rightarrow \infty$
and $m_{\text{sp}}(T)$ is the spontaneous magnetization.
For the two-dimensional Ising model,
$\sigma_\infty (T)$\cite{Onsager44}
and $m_{\text{sp}}(T)$\cite{Yang52} are known exactly.
The length scale on which we would expect a transition from a
single-domain to a multi-domain equilibrium structure
is approximately $L_D$.

The selection of the Ising model is equivalent to requiring
a very large (infinite, in fact) anisotropy constant.
Although magnetic materials used in magnetic recording media
require comparatively large anisotropy constants,\cite{Koester}
the microscopic anisotropy tends to be much smaller than
the exchange energy.
However, the role of the anisotropy is enhanced by coarse-graining.
Simplicity is our main reason for choosing the
two-dimensional Ising model with periodic boundary conditions,
particularly since many
equilibrium properties of the two-dimensional Ising model in
zero field are known exactly\cite{Onsager44,Yang52} and since the
kinetics of metastable decay has been extensively studied
for this model.\cite{RikARCP94}
As a result, our model systems more closely resemble ultrathin
magnetic films with perpendicular magnetization than magnetic grains.
A more realistic simulation of three-dimensional grains
is planned for later study, but we emphasize
that we expect that droplet theory applies to
almost any spin model with high anisotropy.
Accordingly, equations are written in forms appropriate for
arbitrary dimensionality $d$, even though simulations are only
carried out for $d \! = \! 2$.

The relaxation kinetics is simulated by the single-spin-flip
Metropolis dynamic with updates at randomly chosen sites.
A rigorous derivation from microscopic quantum Hamiltonians
of the stochastic Glauber dynamic used in
Monte Carlo simulations of Ising models
has been established in the thermodynamic limit;\cite{Martin77}
both the Glauber and Metropolis algorithms are spatially local
dynamics with non-conserved order parameter
(the dynamic universality class of Model A in
the classification scheme of Hohenberg and
Halperin\cite{Hohenberg77}) and
are therefore expected to differ only in non-universal features.
The Metropolis dynamic is realized by the original
Metropolis algorithm\cite{Metro53} and the
$n$-fold way algorithm.\cite{Bortz75}
(For a discussion on the equivalence of the dynamics of these
algorithms, see Ref.~\onlinecite{NovCIP}.)
The acceptance probability in the Metropolis algorithm
for a proposed flip of the spin at site $\alpha$ from $s_\alpha$
to $-s_\alpha$ is defined as
$W(s_\alpha \! \rightarrow \! -s_\alpha)
	\! = \! \min [1, \exp (-\beta\Delta E_\alpha)]$,
where $\Delta E_\alpha$ is the energy change due to the flip
and $\beta^{-1} \! \equiv \! k_{B}T$ is the temperature
in units of energy.
The $n$-fold way algorithm is similar, but involves the
tabulation of energy classes.  First an energy class is
chosen randomly with the appropriately weighted probability.
A single site is then chosen from within that class with
uniform probability and flipped with probability one.
The number of Metropolis algorithm steps which would be required
to achieve this change is chosen from a geometric probability
distribution,\cite{NovCIP} and the time, measured
in Monte Carlo steps per spin (MCSS), is incremented accordingly.
The $n$-fold way algorithm is more efficient than
the Metropolis algorithm at low temperatures, where the
Metropolis algorithm requires many attempts before a change is
made.

Because we are using single-spin-flip dynamics, the magnetization
can only change by a small amount from one time step to the next.
The dynamical effects of the demagnetizing field thus depend
only on the {\em change} in the magnetic part of ${\cal H}_D$
between adjacent values of the magnetization.  In this way it is
possible to define an {\em effective} magnetic field
\begin{eqnarray}
   H_{\text{eff}}(H,D,m) & \equiv & \frac{\partial}{\partial m}
		\left( Hm - D m^2 \right) \nonumber \\
  \label{eq:Heff}
	& = & H - 2Dm \; .
\end  {eqnarray}
The effective magnetic field is thus site-independent.
This fact makes analytic considerations significantly easier
and is our principal reason for using
Eq.~(\ref{eq:Ham_D}) rather than Eq.~(\ref{eq:Ham_calD}) as the
Hamiltonian.

We study the relaxation of the dimensionless system magnetization
starting from an initial state magnetized opposite to the
applied field [$m(t \! = \! 0) \! = \! +1, H \! < \! 0$].
This approach has often been used in previous studies,
\eg\ in Refs.~\onlinecite{Rik94,Stauffer92}.
For the temperatures employed in this study, the equilibrium
spontaneous magnetizations in zero field are close to unity,
with $0.95 \! < \! m_{\text{sp}} \! < \! 1$.  Since the applied field
is negative (and generally small), the stable magnetization
is 
approximately $m_{\text{st}} \! \approx \! - m_{\text{sp}}$
and the metastable magnetization is
$m_{\text{ms}} \! \approx \! + m_{\text{sp}}$.
We use as an operational definition of the lifetime $\tau$
of the metastable phase the mean first-passage time to a
cutoff magnetization $m \! = \! 0$:
\begin{equation}
	\label{eq-define_tau}
		\tau \equiv \langle t(m \! = \! 0) \rangle \ .
\end  {equation}
It has been observed\cite{Rik94} that the qualitative results
discussed below are not sensitive to the cutoff magnetization
as long as it is sufficiently less than $m_{\text{sp}}$.
Our choice of $m \! = \! 0$ as the cutoff facilitates comparison
with MFM experiments, which are only capable of measuring the
sign of the particle magnetization.

In this paper a numerical subscript indicates the coefficient in
a Taylor expansion in $D$.  For example, a quantity $X$ may be
expanded $X \! = \! X_0 + X_1 D + X_2 D^2 + \ldots$.
There are three exceptions to this rule:
    (1)~The subscripts in Eq.~(\ref{eq:per6b}) refer to an
	iterative process for evaluating a continued fraction.
    (2)~The subscripts on $\Xi_0(T)$ and $\Xi_1(T)$
	[Eq.~(\ref{eq:NucRate})] indicate an expansion in $H^2$
	and are kept for continuity with Ref.~\onlinecite{2dpi}.
    (3)~Dummy variables in the Appendix
	[\eg, Eq.~(\ref{eq:apprVola})]
	may have numerical indices as a matter of convenience.

\section{The Stochastic Region}
\typeout{The Stochastic Region}
\label{sec-stochastic}

It has been shown\cite{Schulman80,Schulman90,Lee95}
that the dynamics of metastable decay
in the standard two-dimensional Ising model
for sufficiently weak applied field
can be semiquantitatively described by a mean-field-like dynamic
in which the free energy is a function only of the system
magnetization.
Under these circumstances switching is abrupt,
with a negligible amount of time being spent in configurations
with magnetizations significantly different from
$m_{\text{ms}}$ or $m_{\text{st}}$.
Switching is also a Poisson process, with
the lifetime of a metastable phase given by the typical
Arrhenius form
\begin{equation}
  \label{eq:Arrhenius}
	\tau \propto \exp \left( \beta \Delta F \right) \; ,
\end  {equation}
where $\Delta F$ is the free-energy barrier that must be crossed
in the decay process, or by a simple generalization of
Eq.~(\ref{eq:Arrhenius}) if more than one equivalent
decay path is present [see Eq.~(\ref{eq:SDLife})].
This phenomenon, in which the entire system behaves as though
it were a single magnetic moment, is known as
superparamagnetism.\cite{Jacobs,Bean59}
The standard deviation of the switching time for an individual
grain is approximately equal to the mean switching time, $\tau$.
Because of the random nature of switching in this region,
it has been called\cite{Tomi92A,Rik94} the ``Stochastic'' region.
The Stochastic region is the union of the ``Coexistence'' region
and the ``Single-Droplet'' region (discussed below).

\begin{figure}
	 \includegraphics{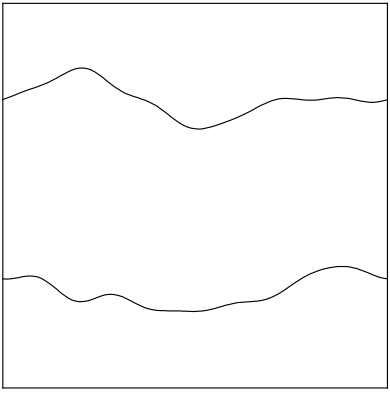}
%
%
	  \noindent
	\caption{
    	 	 \label{fig:Configs}
		Configurations that may occur during the
		reversal process. As in the text, periodic
		boundary conditions are imposed.
		(a)~A sketch of a ``slab'' configuration.
		(b)~A typical realization of a single droplet
			in the process of overtaking the system.
			Grey squares are ``up'' spins and black
			squares are ``down'' spins.
			Here $L \! = \! 60$, $H \! = \! -0.08 J$,
			$D \! = \! 0$, $T \! = \! 0.8 T_c$,
			and $t \! = \! 410$ MCSS.
			[Figure courtesy S.~W.\ Sides.]
		(c)~A typical realization showing the nucleation
			and growth of several droplets in the
			process of switching the magnetization.
			Here $L \! = \! 120$, $H \! = \! -0.2 J$,
			$D \! = \! 0$, $T \! = \! 0.8 T_c$,
			and $t \! = \! 114$ MCSS.
			[Reproduced from Fig.~5(b) of
			Ref.~\protect\onlinecite{2dpi}.]
		}
\end  {figure}

\begin{figure}
	 \includegraphics{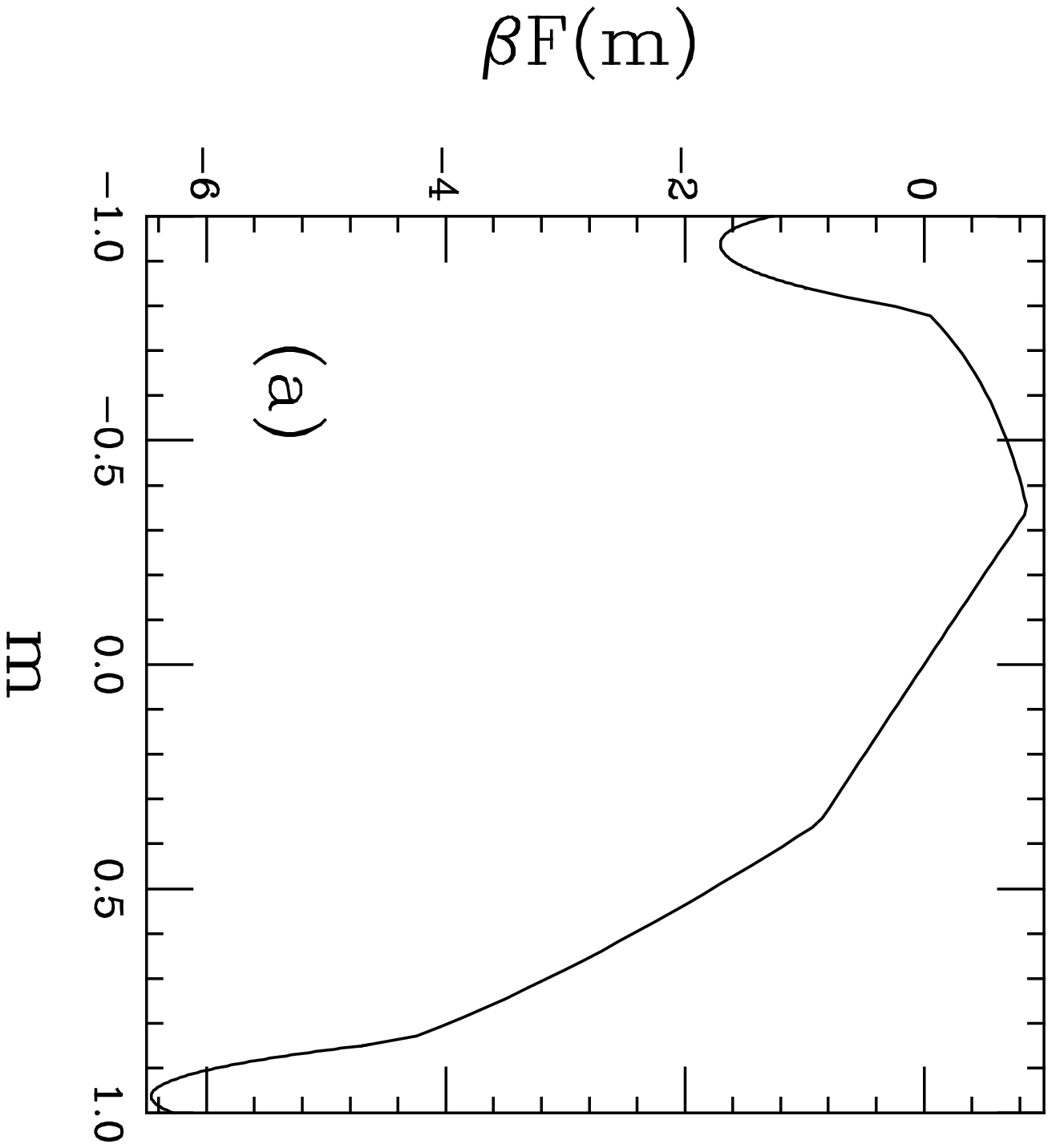}
	 \includegraphics{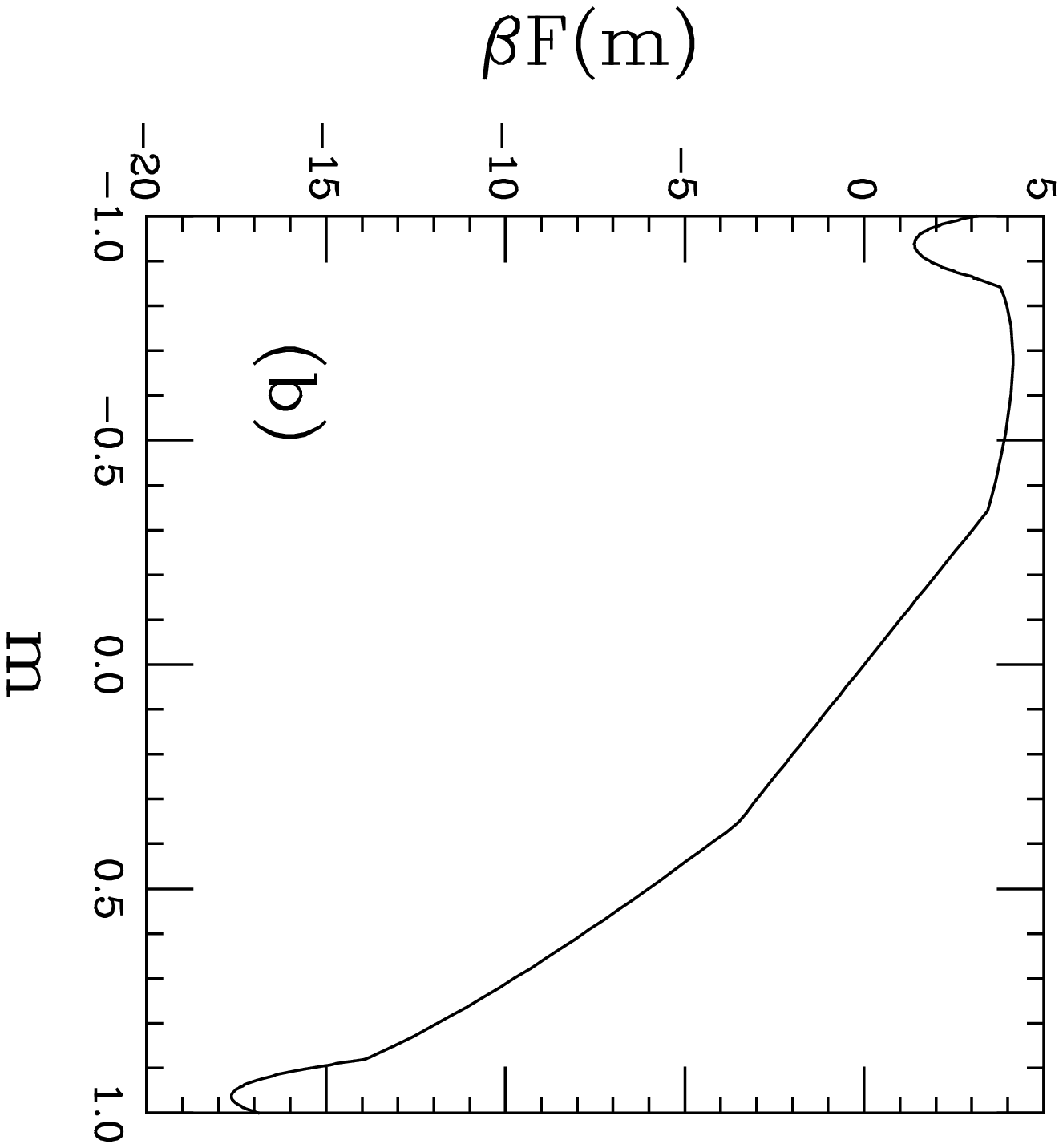}
	 \includegraphics{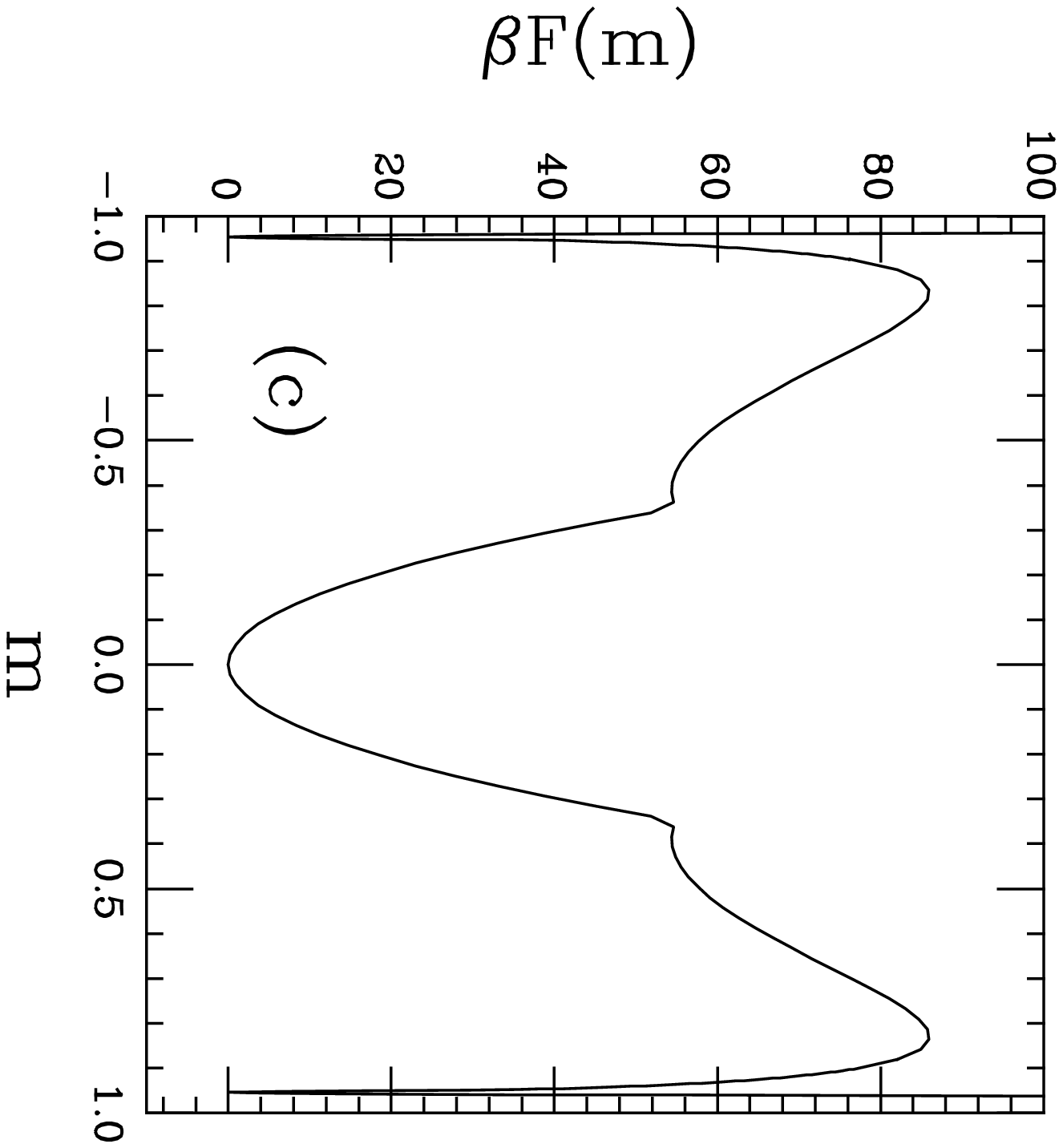}
	 \includegraphics{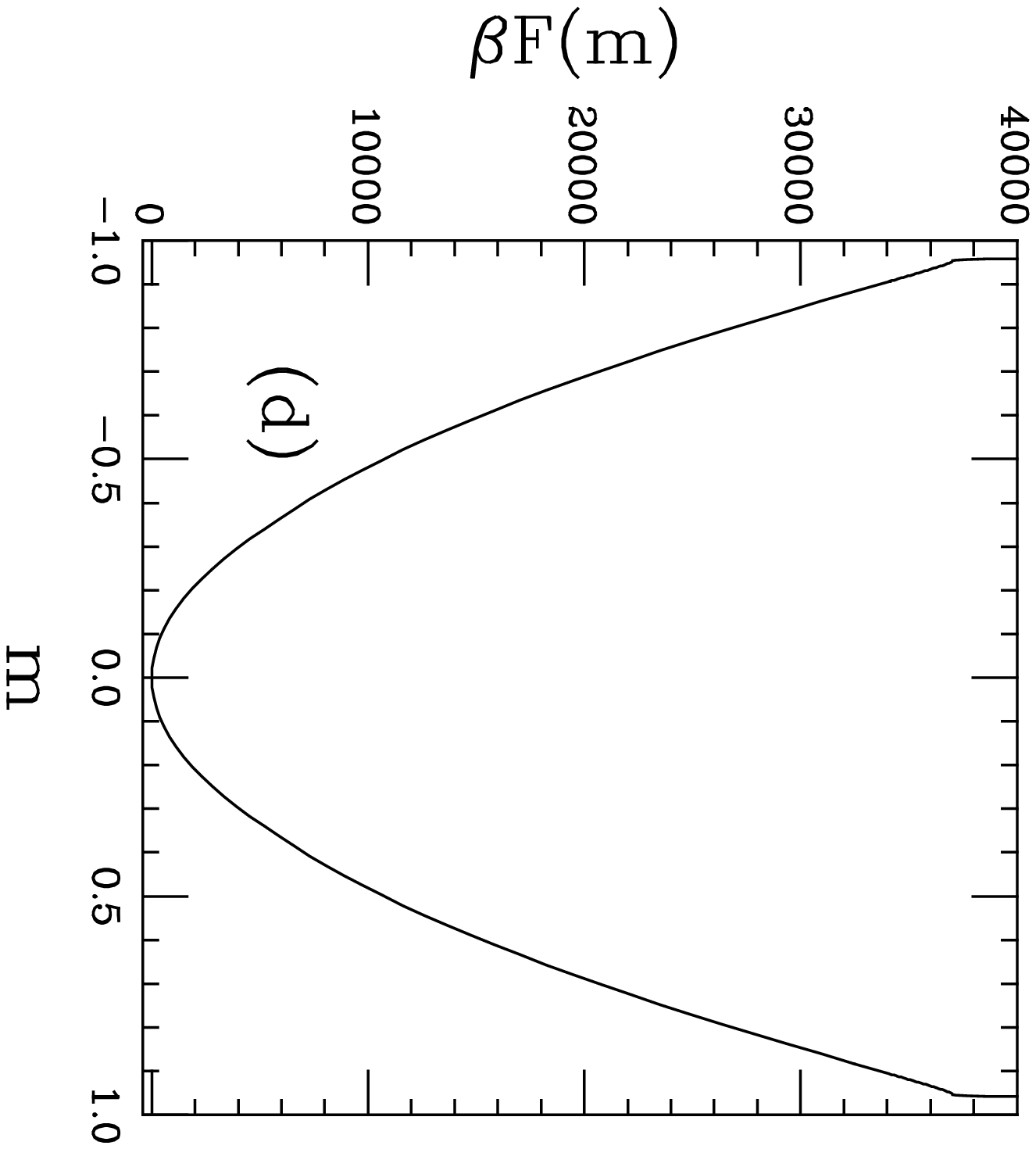}
	  \noindent
	\caption{
    	 	 \label{fig:F}
		The approximate restricted free energy function
		$F(m)$ as determined by
		Eq.~(\protect\ref{eq:Free_m}) with
		$d \! = \! 2$,
		$T \! = \! 0.8 T_c$, and $L_D \! = \! 500$.
		(a)~$L \! = \!    5$, $H \! = \! 0.1 J$.
		(b)~$L \! = \!   10$, $H \! = \! 0.1 J$.
		(c)~$L \! = \!  500$, $H \! = \! 0$.
		(d)~$L \! = \! 5000$, $H \! = \! 0$.
		}
\end  {figure}

In the spirit of Refs.~\onlinecite{Schulman80,Schulman90,Lee95}
we construct an {\em approximate} restricted
free-energy function $F(m)$ for the entire system
and use Eq.~(\ref{eq:Arrhenius}) to illustrate the $H$- and
$D$-dependence of the lifetime:
\begin{equation}
  \label{eq:Free_m}
	F(m) =  L^d D m^2 +
		\min \Bigl\{
		F_{\text{u},+}(m),
		F_{\text{u},-}(m),
		F_{\text{d},+}(m),
		F_{\text{d},-}(m),
		F_{\text{sl}} (m)
		\Bigr\} - F_{\text{sl}}(m=0) \; ,
\end  {equation}
where
$F_{\text{sl}}(m)$ is the free energy of a system composed of two
``slabs'' with magnetizations near $\pm m_{\text{sp}}$
[Fig.~\ref{fig:Configs}(a) illustrates a ``slab'' configuration],
$F_{\text{d},\pm}(m)$ is the free energy of a system with a
single droplet with magnetization near $\mp m_{\text{sp}}$
in a background with magnetization near $\pm m_{\text{sp}}$
[Fig.~\ref{fig:Configs}(b) illustrates a
``single droplet'' configuration], and
$F_{\text{u},\pm}(m)$ is the free energy of a
system in a ``uniform'' phase near $m \! = \! \pm m_{\text{sp}}$.
Figure~\ref{fig:F} illustrates $\beta F(m)$ for $L \! \ll \! L_D$,
$L \! = \! L_D$, and $L \! \gg \! L_D$.

We approximate the free energy of a system
in a ``uniform'' phase by
\begin{equation}
  \label{eq:Free_u}
	F_{\text{u},\pm}(m) \equiv - L^d H m + L^d
		\frac{1}{2} \chi^{-1} (m \mp m_{\text{sp}})^2,
\end  {equation}
where $\chi$ is the equilibrium susceptibility per spin.
Since an exact solution for
the two-dimensional Ising model in a magnetic field has not
yet been found, we use instead an estimate from a series
expansion,\cite{Domb_PTCPv3}
so that for $T \! = \! 0.8 T_c$,
$\chi \! \approx \! 0.05 J^{-1}$.

Minimizing $F_{\text{u},-}(m) \! + \! L^d D m^2$
yields the stable
magnetization $m_{\text{st}}$, which is the location of the
global minimum of $F(m)$ for $L \! < \! L_D$:
\begin{mathletters}
\begin{equation}
  \label{eq:m_st}
	m_{\text{st}} \approx \frac{ -m_{\text{sp}} + H \chi}
		                {     1 + 2 D \chi    }
\end  {equation}
(remember, $H \! < \! 0$).
Likewise, for $L \! \ll \! L_D$ the next-lowest minimum of
$F(m)$ is obtained my minimizing
$F_{\text{u},+}(m) \! + \! L^d D m^2$:
\begin{equation}
  \label{eq:m_ms}
	m_{\text{ms}} \approx \frac{  m_{\text{sp}} + H \chi}
		                {     1 + 2 D \chi    } \; .
\end  {equation}
\end{mathletters}
Equation~(\ref{eq:m_st}) is valid for a wider range of
$H$ than is Eq.~(\ref{eq:m_ms}).
We shall refer to $m_{\text{ms}}$ as the ``metastable magnetization''
and its basin of attraction as the ``metastable phase'', since
{}for systems of interest ($L \! < \! L_D$) the length of time
required for a system initially prepared in the metastable phase
to escape to the stable phase is much longer than any
other timescale.  Note, however, that other, shorter-lived metastable
phases may exist (discussed below).

In cases where the magnetization differs significantly from
$m_{\text{sp}}$ ($-m_{\text{sp}}$), a lower free energy can often be
obtained by segregating the system into a single localized ``droplet''
with magnetization
near $m_{\text{st}}$ ($m_{\text{ms}}$) in a background
with magnetization
near $m_{\text{ms}}$ ($m_{\text{st}}$).\cite{Furukawa82}
Specifically, the droplet free energy is approximated by
\begin{mathletters}
  \label{eq:Free_d}
\begin{equation}
  \label{eq:Free_dp}
	F_{\text{d},+}(m) \equiv
		\Omega \left[
		d \sigma_{\infty} R_{+}^{d-1}
		+
		 ( m_{\text{ms}} - m_{\text{st}} ) H R_{+}^{d}
		\right] - L^d H m_{\text{ms}}
\end  {equation}
and
\begin{equation}
  \label{eq:Free_dm}
	F_{\text{d},-}(m) \equiv
		\Omega \left[
		d \sigma_{\infty} R_{-}^{d-1}
		-
		 ( m_{\text{ms}} - m_{\text{st}} ) H R_{-}^{d}
		\right] - L^d H m_{\text{st}}
\end  {equation}
\end  {mathletters}
subject to $m_{\text{ms}} \! > \! m \! > \! m_{\text{st}}$.
Here
\begin{mathletters}
\begin{equation}
  \label{eq:R+}
	R_{+} = \Omega^{-1/d} L \left(
		\frac{m_{\text{ms}} - m            }
	             {m_{\text{ms}} - m_{\text{st}}}
	                       \right)^{1/d}
\end  {equation}
is the radius of a droplet of ``down'' (stable) spins
in an ``up'' (metastable) background,
\begin{equation}
  \label{eq:R-}
	R_{-} = \Omega^{-1/d} L \left(
		\frac{m             - m_{\text{st}}}
	             {m_{\text{ms}} - m_{\text{st}}}
	                       \right)^{1/d}
\end  {equation}
\end  {mathletters}
is the radius of a droplet of ``up'' (metastable) spins
in a ``down'' (stable) background.
{}For $T \! \not \approx \! 0$, the droplet shape can be found
{}from a Wulff construction.  The quantity
$\Omega$, which gives the volume of the droplet via
$V \! = \! \Omega R^d$, can be found to arbitrary precision
{}for the two-dimensional Ising model by numerically
integrating over
the exactly known surface tension.\cite{Rottman81,Zia82}

Lastly, near $m \! = \! 0$ the circumference of the droplet
becomes larger than twice the cross-section of the system, and the
lowest free energy is obtained by segregating the system
into two slab-like configurations.\cite{Leung90}
The corresponding slab free energy is approximated by
\begin{equation}
  \label{eq:Free_s}
	F_{\text{sl}}(m) \equiv  2L^{d-1} \sigma_{\infty} - L^d Hm.
\end  {equation}
Comparison with Eq.~(\ref{eq:Free_d}) shows that
$F(m) \! = \! F_{\text{sl}}(m)$ for
$m_{\text{ds},+} \! \geq \! m \! \geq \! m_{\text{ds},-}$,
where\cite{Lee95,Leung90}
\begin{mathletters}
\begin{equation}
  \label{eq:m_ds+}
	m_{\text{ds},+} = m_{\text{ms}}
		- (m_{\text{ms}} - m_{\text{st}})\Omega^{-1/(d-1)}
		\left(\frac{2}{d}\right)^{d/(d-1)} > 0
\end  {equation}
and
\begin{equation}
  \label{eq:m_ds-}
	m_{\text{ds},-} = m_{\text{st}}
		+ (m_{\text{ms}} - m_{\text{st}})\Omega^{-1/(d-1)}
		\left(\frac{2}{d}\right)^{d/(d-1)} < 0 \; .
\end  {equation}
\end  {mathletters}

Note that for
$D \! > \! H/(2m_{\text{ds},-})$,
a local minimum of $F(m)$ occurs for a slab configuration at
$m \! = \! H/(2D)$.
{}For $L \! < \! L_D$ it is a metastable phase,
but for $L \! \geq \! L_D$ and $D \! > \! H/m_{\text{st}}$
it is the global minimum of
$F(m)$ and hence the true stable phase (see Fig.~\ref{fig:F}).
Other interesting features can be obtained by solving
$(\rmd /\rmd m)\left[ L^d D m^2 + %
F_{\text{d},\pm}(m) \right] \! = \! 0$,
which can in general be done only numerically. This reveals that
near $L \! = \! L_D$ short-lived metastable phases can exist for
$m \! > \! m_{\text{ds},+}$ or $m \! < \! m_{\text{ds},-}$.

{}For $L \! < \! L_D$, the system enjoys true coexistence
at zero applied field
between two degenerate equilibrium phases with magnetizations
$m_{\text{ms}}$ and $m_{\text{st}}$.
This leads to the identification\cite{Tomi92A,Rik94}
of a ``Coexistence'' (CE) region within which
$F(m_{\text{ms}}) \! \approx \! F(m_{\text{st}})$.
Within the CE region,
the free energy barrier for tunnelling from
the metastable phase to the stable phase
is approximately the same as
the free energy barrier for tunnelling from
the stable phase to the metastable phase,
so the decay process is both stochastic and reversible.
Specifically, for $L \! \ll \! L_D$, the lifetime of the
metastable phase is given by Eq.~(\ref{eq:Arrhenius}) with
$\Delta F \! = \!  F(m_{\text{ds},+}) - F(m_{\text{ms}})$, so
that\cite{Binder81,Binder82,Berg93}
\begin{eqnarray}
	\tau (L,H,T) \approx A(T) \exp
		\Biggl\{ \beta \Biggl[
		2 \sigma_\infty(T) L^{d-1}
		& - & L^d |H|(m_{\text{ms}}   - m_{\text{ds},+})
			\nonumber \\   \label{eq:CELife}
		& - & L^d D  (m_{\text{ms}}^2 - m_{\text{ds},+}^2)
		\Biggr]
		\Biggr\} \; ,
\end  {eqnarray}
where $A(T)$ is a non-universal prefactor.

{}For $L \! \approx \! L_D$, the maximum of $F(m)$ occurs not
at $m_{\text{ds},+}$ but at a larger magnetization corresponding to
a single critical droplet.  The size of this droplet, however,
is strongly dependent on the system size $L$.  This part of
the Coexistence region is further complicated by the
increasing importance of the metastable phase at
$m \! = \! H/(2D)$ and the aforementioned possibility of
metastable phases near $m \! = \! m_{\text{ds},\pm}$.

{}For $H \! \gg \! D$ the maximum of $F(m)$ may correspond to
a critical droplet the size of which is nearly independent of
system size since it is
determined by the {\em applied} field rather than the
{\em demagnetizing} field.
The applied field at which the CE region crosses over into this
``Single Droplet'' (SD) region\cite{Tomi92A,Rik94} is
called\cite{Tomi92A,Rik94} the ``Thermodynamic Spinodal''
($H_{\text{ThSp}}$).
A useful estimate for this crossover is
$(\rmd /\rmd m) \left. \left[ L^d D m^2 +
F_{{\rm d},+}(m)
\right]\right|_{m_{\scriptstyle \text{ds},+}} \! = \! 0$,
yielding\cite{Lee95}
\begin{equation}
	\label{eq:H_THSP}
	|H_{\text{ThSp}}| \approx L^{-1} \Omega^{1/(d-1)}
		\frac{(d-1)\sigma_\infty}{m_{\rm ms} - m_{\rm st}}
		\left( \frac{d}{2} \right)^{1/(d-1)}
		- 2D m_{\rm ds,+} \; .
\end  {equation}
A slightly different estimate was used in
Ref.~\onlinecite{RikARCP94} and Ref.~\onlinecite{2dpi},
but the two estimates are approximately equal.

In the SD region the first critical droplet to nucleate
almost always grows to fill the system before any
other droplet nucleates.
The average time required to nucleate the first droplet
can be estimated from Eq.~(\ref{eq:Arrhenius}), where the
free-energy barrier is determined from
Eq.~(\ref{eq:Free_u}) and Eq.~(\ref{eq:Free_dp}). Since the SD region is
also the region of weak $H$ and $D$, we can obtain a good approximation by
neglecting terms of $O(H_{\text{eff}}\chi)$.
Then the magnetization in the metastable
background is $m_{\text{ms}} \! = \! m_{\text{sp}}$,
and inside the droplet it
is $m_{\text{st}} \! = \! -m_{\text{sp}}$.
In terms of the droplet radius $R$, the difference between the free
energy of a system containing one droplet and that of a uniform
metastable system can then be written as
\begin{equation}
\Delta F(R) = d \Omega \sigma_\infty(T) R^{d-1}
			- 2 m_{\text{sp}} \left(|H|
                    + 2 D m_{\text{sp}}\right) \Omega R^d
			+ 4 D m_{\text{sp}}^2 L^{-d} \Omega^2 R^{2d} \; .
\label{eq:per1}
\end{equation}
Differentiating with respect to $R$, we find the implicit equation
satisfied by
the critical
droplet radius:
\begin{mathletters}
\begin{equation}
	R_{c}(T,H,D) = \frac{(d-1)\sigma_\infty(T)}
		       {2 m_{\text{sp}} |H_{\text{eff},c}(H,D)|} \; ,
\label{eq:per2}
\end{equation}
where
\begin{equation}
|H_{\text{eff},c}(H,D)|
	= |H| \! + 2 D m_{\text{sp}} \left[ 1  -
	2 \Omega \left( R_c/L \right)^d \right]
\label{eq:per3}
\end{equation}
\label{eq:per23}
\end{mathletters}
\noindent
is the effective field evaluated at the magnetization of a
system containing a single, critical droplet.
Note that $H_{\text{eff},c}$
depends on $|H|$ and $D$ explicitly, as well as implicitly through $R_c$.
{}For $D \! = \! 0$, Eq.~(\ref{eq:per23})
reduces to the standard expression for
$R_c$.\cite{RikARCP94}
By inserting $R_c$ into Eq.~(\ref{eq:per1}) one finds that the free-energy
barrier corresponding to the critical droplet is also simply given by a
standard expression,\cite{RikARCP94} in which $|H|$ has been
replaced by $|H_{\text{eff},c}|$:
\begin{equation}
   \beta \Delta F_{\text{SD}}  =
	\frac{\Xi_0(T)}{|H_{\text{eff},c}(H,D)|} \;  ,
\label{eq:per4}
\end{equation}
where\cite{RikARCP94}
\begin{equation}
  \label{eq:def-Xi0}
	\Xi_0 (T) \equiv \beta \Omega
		\left[ \sigma_{\infty} (T)
		\right]^d
		\left(
		  \frac{ d-1 } {2m_{\text{sp}}}
		\right)^{d-1} \; .
\end  {equation}
Note that $\Xi_0(T)$ is completely defined by quantities that for the
two-dimensional Ising model are either known exactly
($\sigma_\infty$ and $m_{\text{sp}}$)\cite{Onsager44,Yang52}
or can be obtained by numerical integration of exactly
known quantities ($\Omega$).\cite{Rottman81,Zia82}

The above results indicate that in order to obtain the nucleation
rate for nonzero $D$, one only needs
to determine $|H_{\text{eff},c}|$. This can easily be done to arbitrary
numerical precision via a rapidly convergent, generalized
continued-fraction expansion as follows.

Let $x \! = \! 2 D m_{\text{sp}}/|H|$ be the
reduced demagnetization field
and $V(x) \! = \! 2 \Omega (R_c/L)^d$ be the volume fraction occupied
by the critical droplet. Then
\begin{equation}
|H_{\text{eff},c}| = |H| \left\{ 1 + x \left[ 1 - V(x) \right] \right\}
                  \equiv |H| \, y(x) \; ,
\label{eq:per5}
\end{equation}
and $V(x)$ is given by the generalized continued-fraction expansion,
\begin{equation}
V(x) = \frac{V_0}
{\left[ 1 + x \left( 1 - \frac{V_0}
{\left[ 1 + x \left( 1 - \frac{V_0}
{\left[ \dots \right]^d}\right) \right]^d}
\right) \right]^d} \; ,
\label{eq:per6}
\end{equation}
where $V_0 \! = \! V(0)$. This expansion can be evaluated to desired
precision by the recursion relation,
\begin{equation}
V_n = \frac{V_0}{\left[ 1 + x \left( 1 - V_{n-1} \right) \right]} \; .
\label{eq:per6b}
\end{equation}
(Here the subscript is proportional to
the order of a rational-function approximation
to the generalized
continued fraction, rather than denoting the order
of a term in a power series in
$D$, as elsewhere in this paper.)

\begin{figure}
\vspace*{2.7in}
	 \includegraphics{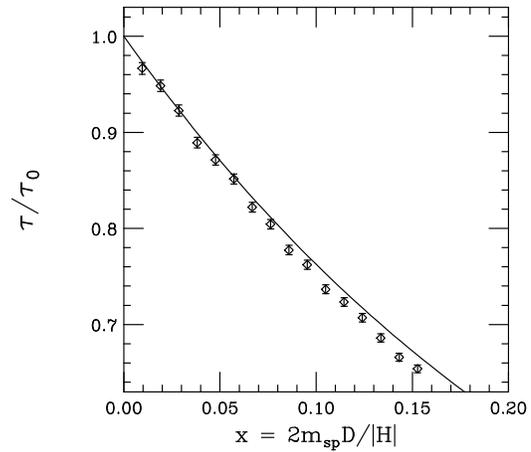}
	  \noindent
	\caption{
    	 	 \label{fig:SD_tau_D}
		The relative lifetime \vs\
		$x \! = \! 2 m_{\text{sp}} D / |H|$
		in the SD region
		as given by Eq.~(\protect\ref{eq:per7}).
		$|H| \! = \! 0.2 J$, $T \! = \! 0.8 T_c$, and
		$L \! = \! 10$.  Each Monte Carlo point
		represents 47~500 decays.
		The quantity $\Xi_1$ was obtained in
		Ref.~\protect\onlinecite{2dpi} from data
		at $D \! = \! 0$ and is {\em not} the result
		of a fit to the $D$-dependence.
		}
\end  {figure}

The lifetime in the SD region can be given in terms of the nucleation
rate per unit volume $I$ by\cite{RikARCP94}
\begin{equation}
  \label{eq:SDLife}
	\tau \approx \left[ L^d I  \right]^{-1} \; .
\end  {equation}
{}For $D \! = \! 0$ and $\chi \! \approx \! 0$,
$I$ has been shown to be given by \cite{RikARCP94}
\begin{equation}
  \label{eq:NucRate}
	I(T,H) \approx B(T) |H|^{K} \exp
	     \left\{ - |H|^{1-d}
		\left[ \Xi_0(T) + \Xi_1(T) H^2 \right]
             \right\} \; ,
\end  {equation}
where $B(T)$ is a non-universal prefactor,
and $K$ is believed to be
3  for the   two-dimensional Ising model and
$-1/3$ for the three-dimensional Ising model.\cite{Rik94}
(The subscripts on $\Xi_0(T)$ and $\Xi_1(T)$
indicate an expansion in $H^2$ rather than in $D$,
and are kept for consistency with
the notation in Ref.~\onlinecite{2dpi}.)
The quantity $\Xi_0 (T)$ is given by Eq.~(\ref{eq:def-Xi0}),
and we determine $\Xi_1(T)$ from a numerical fit
to the $H$-dependence of the lifetime.
Note that the lifetime obtained from Eqs.~(\ref{eq:SDLife}) and
(\ref{eq:NucRate}) is quite similar to the Arrhenius form
with the free-energy barrier given by Eq.~(\ref{eq:per4}).
The only differences are the prefactor $|H|^{-K}$ and
the term $\Xi_1(T) H^2$ in the exponential, which are due to
surface fluctuations on the droplet and to higher-order terms in a
field-theoretical calculation of the free-energy barrier,
respectively.\cite{RikARCP94}
We generalize to the $D \! \neq \! 0$ case by
assuming that the nucleation rate in the SD region
is given by Eq.~(\ref{eq:NucRate}) with $|H|$ replaced by
$|H_{\text{eff},c}(H,D)|$,
as we have already shown in Eq.~(\ref{eq:per4})
for the dominant term in the free-energy barrier,
$\Delta F_{\text{SD}}$. The resulting expression for the relative
lifetime for nonzero $D$ is then given by
\begin{equation}
\frac{\tau(x)}{\tau(0)} = y(x)^{-K}
\exp \left\{
 -  \Xi_0 |H|^{1-d} \left[ 1 \! - \! y(x)^{1-d} \right]
 -  \Xi_1 |H|^{3-d} \left[ 1 \! - \! y(x)^{3-d} \right]
\right\} \; ,
\label{eq:per7}
\end{equation}
where $y(x)$ is defined in Eq.~(\ref{eq:per5}).
This result is shown
in Fig.~\ref{fig:SD_tau_D},
together with MC data,
for $d$=2, $T$=0.8$T_c$, $H$=0.2$J$, and $L$=10.
Except for $\Xi_1$, the parameters needed to
evaluate $\tau(x)/\tau(0)$ are known exactly or numerically exactly
for the two-dimensional Ising model: $V_0$=0.2399(1)
and $\Xi_0$=0.5062(1).
The value of
$\Xi_1$ used in Fig.~\ref{fig:SD_tau_D},
$\Xi_1 \! = \! 9.1(3) J^{-1}$, was obtained in Ref.~\onlinecite{2dpi}
from the $|H|$ dependence of $\tau$ for $D \! = \! 0$ in the
SD region.  Thus the good agreement seen in
Fig.~\ref{fig:SD_tau_D} between
the simulation data and the theoretical prediction is not the
result of a fit to the data, but is determined entirely from
quantities measured with $D \! = \! 0$.

\section{The Multi-Droplet Region}
\typeout{The Multi-Droplet Region}
\label{sec-multidrop}

For sufficiently strong fields or large systems,
decay occurs through many weakly interacting
droplets in the manner described by
Kolmogorov,\cite{Kolmogorov37}
Johnson and Mehl,\cite{JohnsMehl39} and
Avrami.\cite{Avrami}
Such decay is ``deterministic'' in the sense that
the standard deviation of the switching time is
much less than its mean (see Ref.~\onlinecite{2dpi} for details).
The crossover between the SD region and the Multi-Droplet (MD)
region has been called\cite{Tomi92A,Rik94} the
``Dynamic Spinodal'' (DSp).
Since the standard deviation of the lifetime is equal
to its mean in the Stochastic region,
we estimate this crossover by the field $H_{1/2}$ at which
\begin{equation}
	\label{eq:H_DSP}
	\sqrt{ \langle t^2( m \! = \! 0) \rangle - \tau^2 }
		= \frac{\tau}{2} \; .
\end  {equation}
For {\em asymptotically} large $L$,
$H_{\text{DSp}} \! \sim \! (1/\ln L)^{1/(d-1)}$;\cite{Rik94,2dpi}
however, prohibitively large system sizes may be required
before this scaling form is observed.
At a sufficiently high field, nucleation becomes much faster than
growth and the droplet picture breaks down.
The crossover to this ``Strong-Field'' region (SF) has been
called\cite{Tomi92A,Rik94} the ``Mean-Field Spinodal'' (MFSp).
A conservative estimate for this crossover field is obtained by
setting $2 R_c \! = \! 1$:
\begin{equation}
	\label{eq:H_MFSP}
	|H_{\text{MFSp}}| \approx \frac{(d-1) \sigma_\infty(T)}
			    {m_{\text{sp}}} \ .
\end  {equation}
Little is understood quantitatively about the SF region.

In this section we generalize Avrami's Law\cite{Kolmogorov37,%
JohnsMehl39,Avrami} to systems with nonzero $D$.
Avrami's Law
gives the volume fraction of the metastable phase
(or equivalently, the magnetization) for systems in which
droplets nucleate with a constant rate
(per unit volume) $I_0$ and grow at constant velocity $v_0$
without interacting except for overlaps.
The generalization we make below is for a nucleation rate
and velocity that depend on the magnetization, and through it on time.

The time-dependent mean system magnetization $m(t)$
is given by
\begin{equation}
  \label{eq:app-Avrami}
	m(t) = (m_{\text{ms}} - m_{\text{st}}) \rme^{-\Phi (t)}
			+ m_{\text{st}} \; .
\end  {equation}
Here
\begin{equation}
  \label{eq:defPHI}
	\Phi (t)  \equiv  \int_0^t I(t') V(t',t) \rmd t'
\end  {equation}
is the mean volume fraction of droplets (uncorrected for overlap)
and
\begin{equation}
  \label{eq:dropVol}
  V(t_1,t_2) \equiv
	\Omega \left[ \int_{t_1}^{t_2} v(t) \rmd t \right]^{d}
\end  {equation}
is the volume occupied by a droplet which nucleates at time
$t_1$ and grows with a time-dependent radial velocity $v(t)$
until time $t_2$. Here $v(t)$ is the (nonuniversal)
temperature-dependent radial growth velocity of a droplet,
which under an Allen-Cahn
approximation\cite{Lifshitz62,Chan77,Allen79,Filipe95}
is proportional to the effective field in the limit of
large droplets:
\begin{equation}
  \label{eq:def_nu0}
	v(t) \approx \nu \Bigl|H_{\text{eff}} \,
			\biglb(H,D,m(t)\bigrb) \Bigr| \; .
\end  {equation}
The time-dependent nucleation rate is given by
$I(t) \! \equiv \!
I[T,H_{\text{eff}} \, \biglb(H,D,m(t)\bigrb)]$ from
Eq.~(\ref{eq:NucRate}).
Note how this differs from the $D$-dependent nucleation rate
in the SD region. In the SD region, the $D$-dependence of the
nucleation rate comes from the change in system
magnetization from the nucleation of a single critical droplet.
In the MD region, by contrast, we ignore the change in system
magnetization due to the nucleation of a {\em single} droplet
(since $L \! \gg \! R_c$), and the $D$-dependence of the
nucleation rate comes from the change in system magnetization
due to an {\em ensemble} of droplets.

For $D \! = \! 0$, Eq.~(\ref{eq:app-Avrami})
becomes\cite{Kolmogorov37,JohnsMehl39,Avrami}
\begin{equation}
  \label{eq:m0}
	m_0(t) = (m_{\text{ms},0} - m_{\text{st},0})
			\rme^{-\Phi_0 (t)}
			+ m_{\text{st},0} \; ,
\end  {equation}
where
\begin{equation}
  \label{eq:Phi0}
 \Phi_0 (t)  = \ln \left(
		\frac{m_{\text{ms},0} - m_{\text{st},0}}
		    {|m_{\text{st},0}|}                  \right)
		\left( \frac{t}{\tau_0} \right)^{d+1} \; ,
\end  {equation}
so that
$\tau_0$ is the first-passage time to $m \! = \! 0$
[Eq.~(\ref{eq-define_tau})].
Specifically, $\tau_0$ is given
by\cite{Kolmogorov37,JohnsMehl39,Avrami}
\begin{equation}
  \label{eq:tau0}
	\tau_0 = \left[ \frac{I_0 \Omega v_0^d}{(d+1)\ln z_0}
		\right]^{-\frac{1}{d+1}} \; ,
\end  {equation}
where
\begin{equation}
  \label{eq:z0}
	z_0 \equiv \frac{m_{\text{ms},0} - m_{\text{st},0}}
          	                         {|m_{\text{st},0}|}
	\approx  2 \; .
\end  {equation}

\begin{figure}
\vspace*{2.7in}
	 \includegraphics{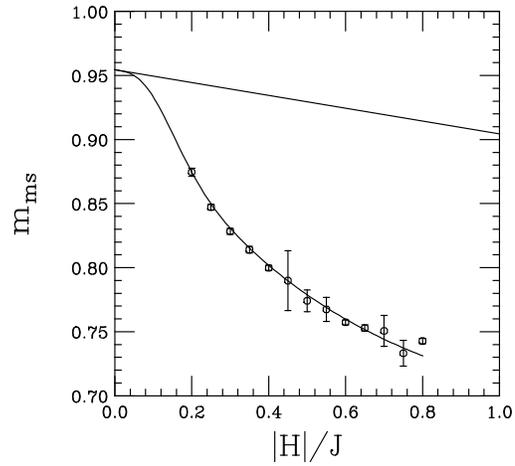}
	  \noindent
	\caption{
    	 	 \label{fig:mms_ramos}
		The metastable magnetization \vs\ $H$
		in the MD region.  The data points were
		estimated by Ramos {\em et al.}\protect\cite{Ramos95}
		by extrapolating
		back to $t \! = \! 0$ assuming
		Avrami's Law.
		The smooth curve is a useful estimate, but
		the correct form of $m_{\text{ms}}$ \vs\ $H$ is
		not known except near $H \! = \! 0$, where
		Eq.~(\protect\ref{eq:m_ms}) applies.
		The straight line indicates the approximation
		Eq.~(\protect\ref{eq:m_ms}).
		}
\end  {figure}

In Eq.~(\ref{eq:z0}) and elsewhere in this section, the estimate
for the metastable magnetization given by Eq.~(\ref{eq:m_ms})
does not suffice. Ramos {\em et al.}\ have estimated
$m_{\text{ms}} (H)$ by
extrapolating $m(t)$ back to $t \! = \! 0$ assuming that
$m(t)$ is correctly described by Avrami's Law.\cite{Ramos95}
(Since the initial condition is $m_0 \! = \! 1$ rather than
$m_0 \! = \! m_{\text{ms}}$, the earliest times must be
discarded.)
These estimates are
shown in Fig.~\ref{fig:mms_ramos}.  We fit a smooth curve through the
data, insisting that $m_{\text{ms}}(0) \! \equiv \! m_{\text{sp}}$
and $(\rmd / \rmd H)m_{\text{ms}}|_{H = 0} \! = \! \chi$.
The smooth curve allows us to estimate the change in
$m_{\rm ms}$ due to $D$, but the correct form of $m_{\text{ms}} (H)$
is not known.

{}For $D \! = \! 0$, the standard deviation of the
time-dependent magnetization
has been shown to vanish with increasing system size as
$L^{-d/2}$,\cite{2dpi,Sekimoto86} a feature which is shared by
the more general case of $D \! \geq \! 0$.  In fact, realizations
in which the magnetization chances to decay more rapidly than
average will experience a weaker effective field
[Eq.~(\ref{eq:Heff})], and realizations
in which the magnetization decays more slowly than
average will experience a stronger effective field.
These effects combine to cause systems with $D \! > \! 0$ to
have even smaller standard deviations in their
time-dependent magnetizations
than corresponding systems with $D \! = \! 0$.

{}To first order in $D$, the effective magnetic field [from
Eq.~(\ref{eq:Heff})] is given by
\begin{equation}
  \label{eq:H_eff1}
	H_{\text{eff}} \, \biglb(H,D,m(t)\bigrb)
	\approx H - 2 D m_0(t)
\end  {equation}
since any $D$-dependent terms in $m(t)$ will lead to only
higher-order corrections.
We will expand $\Phi (t)$ to first order in $D$, so that we
can use the known value of $m_0 (t)$ instead of the unknown
value $m(t)$ on the right-hand side of Eq.~(\ref{eq:defPHI}).
In order to perform the expansion correctly, we must expand
$I(t)$ and $V(t_1,t_2)$ to first order in $D$.
Specifically, the total volume fraction
(uncorrected for overlap) of droplets after
a time $t$ is given by
\begin{equation}
  \label{eq:apprPHI_a}
	\Phi (t) \approx  \Phi_0 (t) + \Phi_1 (t) D
\end  {equation}
where
\begin{equation}
  \label{eq:defPHI_1}
	\Phi_1 (t) \equiv \Phi_V (t) + \Phi_I (t) \; ,
\end  {equation}
\begin{equation}
  \label{eq:defPHI_V}
	\Phi_V (t)  \equiv  \int_0^t I_0 V_1(t',t) \rmd t' \; ,
\end  {equation}
and
\begin{equation}
  \label{eq:defPHI_I}
	\Phi_I (t) \equiv \int_0^t I_1(t') V_0 (t',t) \rmd t' \; .
\end  {equation}

Straightforward but cumbersome mathematics shows (see the Appendix)
\begin{eqnarray}
   \Phi_V (t)
      & \approx & 2 (d) |H|^{-1} |m_{\text{st},0}|
	\Biggl\{ - \Phi_0(t)
	+ \frac{z_0}{d} \left[ \Phi_0 (t) \right]^{d/(d+1)}
		\gamma \left[ \frac{1}{d+1}, \Phi_0 (t) \right]
	\nonumber \\ & & \label{eq:PHIV_result} +
	\frac{z_0}{d+1} \sum_{k=0}^{d-1} \pmatrix{d-1 \cr k}
	(-1)^{d-k} \left[ \Phi_0 (t) \right]^{k/(d+1)}
	{\cal A}\left[ \frac{-(k+1)}{d+1}, \frac{1}{d+1},
		\Phi_0(t) \right]
	\Biggr\}
\end  {eqnarray}
and
\begin{eqnarray}
\Phi_I (t) \approx \lambda(H) |m_{\text{st},0}| & \Bigglb( &
		z_0 \sum_{k=0}^d \left\{
		\pmatrix{d \cr k} (-1)^{d-k}
		\left[ \Phi_0 (t) \right]^{k/(d+1)}
		\gamma \left[ 1 - \frac{k}{d+1}, \Phi_0 (t) \right]
		\right\}
		\nonumber \\ & & \label{eq:PHII_result}
		- \Phi_0 (t)
		\Biggrb) \; ,
\end  {eqnarray}
where
\begin{equation}
  \label{eq:lambda}
	\lambda(H) \equiv 2 \left\{ \frac{K}{|H|} +
		|H|^{-d} \left[ (d-1) \, \Xi_{0}
		 + (d-3) \, \Xi_{1} H^{2} \right] \right\} \; ,
\end  {equation}
comes from differentiating Eq.~(\ref{eq:NucRate}) with respect
to $H$,
$\gamma$ denotes the incomplete gamma function
\begin{equation}
  \label{eq:incgama}
  \gamma (a,x) \equiv  \int_0^x y^{a-1} \rme^{-y} \rmd y \; ,
\end  {equation}
and ${\cal A}$ is given by
\begin{equation}
	\label{eq:defint_A1}
 {\cal A}(a,b,x) \equiv \int_0^x y^a \gamma (b,y) \rmd y \; .
\end  {equation}
Both $\gamma$ and ${\cal A}$ are easily evaluated by
Taylor series.

We can insert $\Phi_V (t)$ and $\Phi_I (t)$
into Eq.~(\ref{eq:defPHI_1}) to find $\Phi_1 (t)$
and then use Eq.~(\ref{eq:app-Avrami}) to evaluate
\begin{equation}
  \label{eq:m1}
	m_1 (t) =  (m_{\text{ms},1} - m_{\text{st},1})
			\rme^{-\Phi_0 (t)}
			+ m_{\text{st},1}
		- (m_{\text{ms},0} - m_{\text{st},0})
			\rme^{-\Phi_0 (t)} \Phi_1 (t) \; .
\end  {equation}

Finding the second-order terms in $D$ proceeds along parallel lines.
Although it is possible to find an analytic expression for
$m_2(t)$, this expression is tedious to derive and unenlightening.
Furthermore, enough approximations have already been introduced to
make the significance of an analytic expression for $m_2(t)$
suspect.  Consequently, we estimate $m(t)$ by
integrating Eq.~(\ref{eq:defPHI}) numerically with the
effective field $H - 2D [m_0(t) + Dm_1(t)]$.  The resulting
estimate we denote $m_{\int}(t)$,
and it should be approximately correct to $O(D^2)$.
If necessary, this process can be iterated to find successively
better approximations for $m(t)$.

\begin{figure}
\vspace*{2.7in}
	 \includegraphics{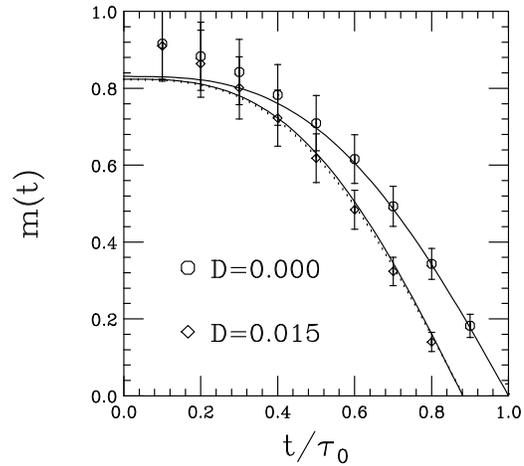}
	  \noindent
	\caption{
    	 	 \label{fig:MD_m_vs_t}
		The magnetization \vs\ time in the MD region
		as given by Eq.~(\protect\ref{eq:m0})
		and Eq.~(\protect\ref{eq:m1}).
		$H \! = \! 0.3 J$, $T \! = \! 0.8 T_c$, and
		$L \! = \! 50$.
		The two values of
		$D$ displayed are $D \! = \! 0$ and
		$D \! = \! 0.015$.
		The solid curves are
		$m_0(t) \! + \! Dm_1(t)$;
		the dotted curve (hardly distinguishable on
		the scale of this figure) is $m_{\int}(t)$.
		Each Monte Carlo point
		represents 100 decays.
		}
\end  {figure}

Figure~\ref{fig:MD_m_vs_t} shows the time-dependence of the
the magnetization, both as approximated above and as simulated
by Monte Carlo.  For $\Xi_1$ we have used $\Xi_1 \! = \! 3.0(3)$,
as determined
in Ref.~\onlinecite{2dpi} from the $|H|$ dependence of $\tau$ for
$D \! = \! 0$ in the MD region.
Note that there is good agreement
between the simulation results and the approximation
after an initial relaxation into the metastable
phase, and that the modification to $m(t)$
resulting from higher-order terms is minor.

In order to calculate the effect of $D$ on the lifetime, we
start with $m(\tau) \! \equiv \! 0$ and expand both $m$ and
$\tau$ in $D$.  Collecting terms and discarding all terms
of higher order than $D^2$, we find
\begin{eqnarray}
0 & = & \{m_0(\tau_0) \} +
	\left\{ m_1(\tau_0)
	+ \tau_1 \left. \frac{\rmd m_0}{\rmd t} \right|_{\tau_0}
	\right\} D
\nonumber \\ & & \label{eq:findtauD}
	+ \left\{ m_2(\tau_0)
	+ \tau_1 \left. \frac{\rmd m_1}{\rmd t} \right|_{\tau_0}
	+ \tau_2 \left. \frac{\rmd m_0}{\rmd t} \right|_{\tau_0}
	+ \frac{1}{2} \tau_1^2
	 \left. \frac{\rmd^2 m_0}{\rmd t^2} \right|_{\tau_0}
	\right\} D^2 \; .
\end  {eqnarray}
Since the quantities in the braces are independent of $D$ and
Eq.~(\ref{eq:findtauD}) is true for all small $D$, by necessity
$m_0(\tau_0) \! = \! 0$,
\begin{equation}
  \label{eq:findtau1}
	\tau_1 = - m_1(\tau_0) \left[ \left.
			\frac{\rmd m_0}{\rmd t}
			\right|_{\tau_0} \right]^{-1} \; ,
\end  {equation}
and
\begin{equation}
  \label{eq:findtau2}
	\tau_2 = - \left[ m_2(\tau_0)
	+ \tau_1 \left. \frac{\rmd m_1}{\rmd t} \right|_{\tau_0}
	+ \frac{1}{2} \tau_1^2
	 \left. \frac{\rmd^2 m_0}{\rmd t^2} \right|_{\tau_0}
	\right] \left[ \left.
	\frac{\rmd m_0}{\rmd t} \right|_{\tau_0} \right]^{-1} \; .
\end  {equation}

Equation~(\ref{eq:findtau1}) is readily evaluated because
$\Phi_0 (t)$ is a simple function. It is likewise simple
to calculate $(\rmd^2 m_0 / \rmd t^2)|_{\tau_0}$ for use in
Eq.~(\ref{eq:findtau2}).  We have not actually solved for
$m_2(t)$, but for small $D$ we can use
\begin{equation}
  \label{eq:approxm2}
	m_2(t) \approx D^{-2} \left\{ m_{\int}(t)
	        - \left[ m_0(t) + D m_1(t) \right] \right\} \; .
\end  {equation}
Finally, $(\rmd m_1 / \rmd t)|_{\tau_0}$ can be evaluated by
differentiating Eq.~(\ref{eq:PHIV_result}) and
Eq.~(\ref{eq:PHII_result}) with respect to $t$:
\begin{eqnarray}
   \left. \frac{\rmd}{\rmd t}\Phi_V (t) \right|_{t=\tau_0}
      & \approx & 2 (d) |H|^{-1} |m_{\text{st},0}|
	\frac{(d+1)\ln z_0 }{\tau_0}
	\nonumber \\ & &
	\times \Biggl\{ -1 + \frac{1}{d}
	\frac{z_0}{d+1} \left( \ln z_0 \right)^{-1/(d+1)}
		\gamma \left( \frac{1}{d+1}, \ln z_0 \right)
	\nonumber \\ & &
	+ \frac{z_0}{(d+1)^2} \sum_{k=0}^{d-1} \pmatrix{d-1 \cr k}
	(-1)^{d-k} k \left( \ln z_0 \right)^{(k-d-1)/(d+1)}
	\nonumber \\ & & \label{eq:dPHIV_dt}
	{\cal A}\left( \frac{-(k+1)}{d+1}, \frac{1}{d+1},
		\ln z_0 \right)
	\Biggr\}
\end  {eqnarray}
and
\begin{eqnarray}
   \left. \frac{\rmd}{\rmd t}\Phi_I (t) \right|_{t=\tau_0}
& \approx & \lambda(H) |m_{\text{st},0}| \frac{(d+1) \ln z_0}{\tau_0}
		\Bigglb( - 1 +
		\frac{z_0}{d+1} \sum_{k=0}^d \Biggl\{
		\pmatrix{d \cr k} (-1)^{d-k} k
		\nonumber \\ & & \label{eq:dPHII_dt}
		\times
		\left( \ln z_0 \right)^{(k-d-1)/(d+1)}
		\gamma \left( 1 - \frac{k}{d+1}, \ln z_0 \right)
		\Biggr\}
		\Biggrb) \; .
\end  {eqnarray}
This yields $(\rmd \Phi_1 / \rmd t)|_{\tau_0}$ from
Eq.~(\ref{eq:defPHI_1}).  Once this is known, differentiating
Eq.~(\ref{eq:m1}) is trivial, and $\tau_2$ can
easily be evaluated.

\begin{figure}
\vspace*{2.7in}
	 \includegraphics{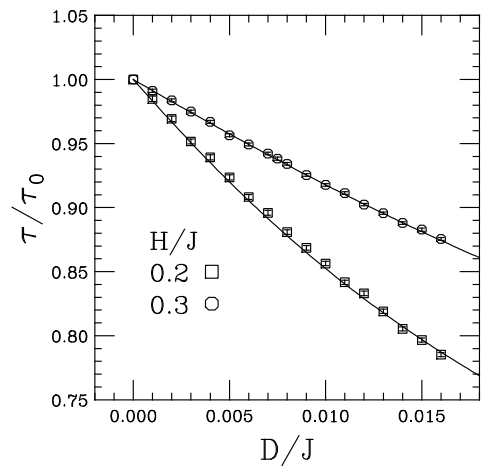}
	  \noindent
	\caption{
    	 	 \label{fig:MD_tau_D}
		The lifetime \vs\ $D$ in the MD region for
		$T \! = \! 0.8 T_c$  with
		$H \! = \! 0.2 J$ and $H \! = \! 0.3 J$.
		The solid curves represent the theoretical
		predictions given by combining
		Eq.~(\protect\ref{eq:findtau1})
		and Eq.~(\protect\ref{eq:findtau2}) and
		are {\em not} the result of a fit to the
		$D$-dependence. Each Monte Carlo point
		represents at least 5~000 decays in a system
		of size $L \! = \! 100$.
			[Reproduced from Fig.~4 of
			Ref.~\protect\onlinecite{MMM95}.]
		}
\end  {figure}

Figure~\ref{fig:MD_tau_D} shows $\tau$ \vs\ $D$ for two different
values of $H$.  The agreement between the theoretical
curves and the Monte Carlo data is again excellent.  Once again,
the theoretical curves are not fits to the simulation data, but
use only parameters determined for $D \! = \! 0$, namely,
$m_{\text{ms}}(H)$\cite{Ramos95}  and $\Xi_1(T)$.\cite{2dpi}

\section{Discussion}
\typeout{Discussion}
\label{sec-discuss}

Due to the importance of magnetic recording technologies in
modern society, magnetic relaxation has been a subject of study
for many years.  However, even the equilibrium thermodynamics
of magnetic materials is very difficult to predict from first
principles and generally has to be approximated from simpler
models (see, \eg, Ref.~\onlinecite{Mills_UMSv1}).
As a result, the most popular method for theoretical investigation
of magnetization reversal involves setting up and solving
differential equations on a lattice obtained by
course-graining over the microscopic crystal lattice.  This
method, known as micromagnetics,\cite{WFBrown}
often gives very good results, particularly for equilibrium
studies or for multi-domain particles.  However, micromagnetic
calculations take thermal effects into account only crudely.

An alternative method is to treat the statistical mechanics
carefully, making simplifications to the model until
it can be well understood.  This is the approach we take.
In Ref.~\onlinecite{2dpi} we showed that both
the switching field and the probability that the magnetization
is greater than zero, calculated from
Monte Carlo simulations of
the two-dimensional Ising model, are qualitatively
similar to the same quantities measured in isolated, well
characterized single-domain ferromagnets by techniques such as MFM.
Since statistical-mechanical droplet theory successfully
explains the Ising model simulations, it is plausible that
droplet theory could also be applied to the experimental particles.

In this article we consider the effect of the magnetic dipole-dipole
interaction, which was neglected in Ref.~\onlinecite{2dpi}.
By treating the dipole-dipole interaction in a mean-field
approximation, we are able to calculate droplet-theory predictions
for the lifetime for systems in which magnetic decay occurs by
means of a single droplet (Fig.~\ref{fig:SD_tau_D}).
We also obtain both
the time-dependent magnetization (Fig.~\ref{fig:MD_m_vs_t})
and the lifetime (Fig.~\ref{fig:MD_tau_D}) for
systems in which magnetic decay occurs through the
action of many droplets.
In all of these figures, all parameters were determined by
measurements at $D \! = \! 0$, so the excellent agreements
are not the result of curve fitting.

It should be pointed out that the droplet-theory predictions
made in both the Single-Droplet and Multi-Droplet regions are
large-droplet approximations.  Since a critical droplet in the
two-dimensional Ising model at $T \! = \! 0.8 T_c$ and
$|H| \! = \! 0.3 J$ consists of approximately six overturned
spins, it is quite remarkable that these expressions
give the good approximations they do.

In Ref.~\onlinecite{Kirby94}, Kirby {\it et al.}\
use the two-dimensional Ising model with mean-field
dipole-dipole interactions to simulate Dy/Fe ultrathin
films which they have observed experimentally, obtaining
good agreement.  The analytic results from
Sec.~\ref{sec-multidrop} are
directly applicable to such films, although different values
of $T$, $H$, and $D$ must be chosen to make the comparison.
However, care should be used in applying the results.
Specifically, if $|D m_{\text{sp}}/H|$ is not small, more numerical
iterations of the type described in Sec.~\ref{sec-multidrop}
will be necessary, and if
$|H_{\text{eff}}| \! \gtrsim \! |H_{\text{MFSp}}|$
[Eq.~(\ref{eq:H_MFSP})], droplet theory may not be applicable.

It is interesting to note that the mean-field demagnetizing field
we have used does not change the qualitative behavior of the
switching field as a function of system size from that which
was studied in Ref.~\onlinecite{2dpi} and sketched in
Fig.~\ref{fig:HsRoad}.  Although the values of the switching
field are reduced, a peak in the switching field still occurs
near the thermodynamic spinodal, as can be seen by
comparing Eq.~(\ref{eq:CELife}) and Eq.~(\ref{eq:SDLife}).
Furthermore, the switching field remains roughly independent of
$L$ in the MD region.  What is noticeably absent is any
feature at $L \! = \! L_D$; the switching field shows features
due to transitions in the dynamics, but not transitions in
the statics.

Even with the addition of the demagnetizing field, the
Ising model remains too crude a model for magnetism to
describe quantitatively real magnetic materials, except
perhaps for some ultrathin films as noted above.
Heterogenous nucleation at boundaries, quenched disorder, and
more realistic anisotropies are therefore planned
as the subject of future studies.

\acknowledgements

The authors wish to thank B.~M.~Gorman and
R.~A.\ Ramos for useful discussions and for comments on the manuscript.
We also wish to thank S.~W.\ Sides for Fig.~\ref{fig:Configs}(b), and
R.~A.\ Ramos for providing the data on which
Fig.~\ref{fig:mms_ramos} is based prior to publication.
During the final stages of this work, H.~L.~R.\ and
P.~A.~R.\ enjoyed the hospitality and support of the
Ris{\o} National Laboratory and McGill University, respectively.
This research was supported in part by the Florida State University
Center for Materials Research and Technology, by the FSU
Supercomputer Computations Research Institute, which is partially
funded by the U.~S.\ Department of Energy through Contract No.\
DE-FC05-85ER25000,
and by the National Science Foundation through Grants
No.\ DMR-9315969 and DMR-9520325.
Computing resources at the
National Energy Research Supercomputer Center were made available by the
U.~S. Department of Energy.


\appendix
\section*{Avrami's Law for $D \! \geq \! 0$}
\label{app_details}

In this Appendix we give some of the steps that have been omitted
for clarity in Sec.~\ref{sec-multidrop}.
Beginning with the Allen-Cahn approximation\cite{Lifshitz62,Chan77,%
Allen79,Filipe95} for the radial
growth velocity [Eq.~(\ref{eq:def_nu0})]
and with the effective magnetic field to $O(D)$
[Eq.~(\ref{eq:Heff})], we find
\begin{equation}
  \label{eq:AllCahn}
	v(t) \approx v_0
		\left[ 1 + \frac{2D}{|H|} m_0(t)
		\right] \; .
\end  {equation}
Substituting Eq.~(\ref{eq:AllCahn})
into Eq.~(\ref{eq:dropVol}), we find
\begin{eqnarray}
  \label{eq:apprVola}
V(t_1,t_2)
& \approx & \Omega v_0^d \left\{ \int_{t_1}^{t_2}
		\left[ 1 + \frac{2D}{|H|} m_0(t)
		\right] \rmd t \right\}^{d}
\nonumber \\  \label{eq:apprVolb} & \approx &
	\Omega v_0^d \left[
		\left( t_2 - t_1 \right)^d
		+  \pmatrix{d \cr 1} \frac{2D}{|H|}
		\left( t_2 - t_1 \right)^{d-1}
		\int_{t_1}^{t_2} m_0(t)  \rmd t \right] \; .
\end  {eqnarray}
This enables us to make the identifications
\begin{equation}
    \label{eq:Vol0}
V_0(t_1,t_2) = \Omega v_0^d (t_2 - t_1)^d
\end  {equation}
and
\begin{equation}
    \label{eq:Vol1}
V_1(t_1,t_2) = 2 \Omega v_0^d
		\pmatrix{d \cr 1} |H|^{-1}
		\left( t_2 - t_1 \right)^{d-1}
		\int_{t_1}^{t_2} m_0(t) \rmd t \; .
\end  {equation}

A Taylor expansion of the nucleation rate
[Eq.~(\ref{eq:NucRate})]
\begin{eqnarray}
  \label{eq:ITaylora}
     I(t)
\approx
	I_0 + \left. \frac{\rmd I}{\rmd D} \right|_0 D
& = &
 	I_0 + \left. \frac{\rmd I}{\rmd |H_{\text{eff}}|} \right|_0
              \left. \frac{\rmd |H_{\text{eff}}|}{\rmd D} \right|_0 D
             \nonumber \\
& = & \label{eq:ITaylorc}
	I_0 + \left. \frac{\rmd I}{\rmd |H_{\text{eff}}|} \right|_0
              \Bigl[ 2 m_0(t) \Bigr] D
\end  {eqnarray}
yields
\begin{equation}
  \label{eq:I1}
	I_1 (t) = \lambda(H) I_0 m_0(t) \; ,
\end  {equation}
where $\lambda(H)$ is given by Eq.~(\ref{eq:lambda}).

The evaluation of $\Phi_I(t)$ and $\Phi_V(t)$ can be followed
more easily if the reader keeps in mind three basic ``tricks'':
we apply of the binomial theorem,
\begin{equation}
  \label{eq:binomthrm}
(t-t')^{n}  = \sum_{k=0}^{n} t^k (-t')^{n-k} \pmatrix{n \cr k} \; ,
\end  {equation}
we use Eq.~(\ref{eq:tau0}) to simplify expressions,
and we make changes of variables of the form
$x \! = \! \Phi_0(t')$.
These lead to expressions such as
\begin{equation}
  \label{eq:changevar}
(d+1)\left(\frac{\ln z_0}{\tau_0^{d+1}}\right)^{(n+1)/(d+1)}
t^{k} (t')^{n-k} \rmd t'
 =  [\Phi_0(t)]^{k/(d+1)} x^{(n-k-d)/(d+1)} \rmd x \; .
\end  {equation}

Using Eq.~(\ref{eq:m0}) in Eq.~(\ref{eq:Vol1})
\begin{eqnarray}
 V_1(t_1,t_2) & \approx & 2 \Omega v_0^d \pmatrix{d \cr 1} |H|^{-1}
		\left( t_2 - t_1 \right)^{d-1} |m_{\text{st},0}|
		\nonumber \\ \label{eq:Vol1_1} & & \times
		\int_{t_1}^{t_2} \left\{
		z_0 \exp \left[ - \ln (z_0) \left(
			\frac{t}{\tau_0} \right)^{d+1} \right] - 1
	\right\}  \rmd t \nonumber \\
& = &
	2 \pmatrix{d \cr 1} |H|^{-1} |m_{\text{st},0}|
		\Biggl[ -V_0(t_1,t_2)
		+ \Omega v_0^d \frac{z_0}{d+1}
		(\ln z_0)^{-1/(d+1)}
		(t_2-t_1)^{d-1} \tau_0
		\nonumber \\ & & \label{eq:apprVol4}
		\times
		\int_{\Phi_0(t_1)}^{\Phi_0(t_2)}
		\rme^{-x} x^{\frac{1}{d+1} - 1} \rmd x
		\Biggr]
\nonumber \\ & = &
		2 \pmatrix{d \cr 1} |H|^{-1} |m_{\text{st},0}|
		\Biggl( -V_0(t_1,t_2)
		+ \Omega v_0^d \frac{z_0}{d+1}
		(\ln z_0)^{-1/(d+1)}
		(t_2-t_1)^{d-1} \tau_0
		\nonumber \\ & & \label{eq:apprVol5}
		\times
		\left\{
		\gamma \left[ \frac{1}{d+1}, \Phi_0 (t_2) \right] -
		\gamma \left[ \frac{1}{d+1}, \Phi_0 (t_1) \right]
		       \right\} \Biggr) \; .
\end  {eqnarray}

Using Eq.~(\ref{eq:apprVol5}) in Eq.~(\ref{eq:defPHI_V}),
\begin{eqnarray}
	\Phi_V (t)
& \approx & 2 \pmatrix{d \cr 1} |H|^{-1} |m_{\text{st},0}|
	\Bigglb( - \Phi_0(t)
	+ I_0 \Omega v_0^d \frac{z_0}{d+1}
	(\ln z_0)^{-1/(d+1)} \tau_0
	\nonumber \\ & & \label{eq:apprPHIV_1}
	\times
	\left\{
	\frac{t^d}{d}
	    \gamma \left[ \frac{1}{d+1}, \Phi_0 (t) \right] -
	\int_0^t (t-t')^{d-1}
		\gamma \left[ \frac{1}{d+1}, \Phi_0 (t') \right]
		\rmd t'
	\right\} \Biggrb) \nonumber \\
& = &
	2 \pmatrix{d \cr 1} |H|^{-1} |m_{\text{st},0}|
	\Bigglb( - \Phi_0(t)
	+ z_0
	(\ln z_0)^{d/(d+1)} \tau_0^{-d}
	\nonumber \\ & & \label{eq:apprPHIV_2} \times
	\left\{
	\frac{t^d}{d}
	    \gamma \left[ \frac{1}{d+1}, \Phi_0 (t) \right] -
	\int_0^t \sum_{k=0}^{d-1} \pmatrix{d-1 \cr k}
		t^k (-t')^{d-1-k}
		\gamma \left[ \frac{1}{d+1}, \Phi_0 (t') \right]
		\rmd t'
	\right\} \Biggrb) \nonumber \\
& = &
	2 \pmatrix{d \cr 1} |H|^{-1} |m_{\text{st},0}|
	\Biggl\{ - \Phi_0(t)
	+ \frac{z_0}{d} \left[ \Phi_0 (t) \right]^{d/(d+1)}
		\gamma \left[ \frac{1}{d+1}, \Phi_0 (t) \right]
	\nonumber \\ & & \label{eq:apprPHIV_3} +
	\frac{z_0}{d+1} \sum_{k=0}^{d-1} \pmatrix{d-1 \cr k}
	(-1)^{d-k} \left[ \Phi_0 (t) \right]^{k/(d+1)}
	{\cal A}\left( \frac{-(k+1)}{d+1}, \frac{1}{d+1},
		\Phi_0(t) \right)
	\Biggr\} \; .
\end  {eqnarray}

It is somewhat easier to evaluate $\Phi_I (t)$.  From
Eq.~(\ref{eq:defPHI_I}) and Eq.~(\ref{eq:I1}),
\begin{eqnarray}
  \label{eq:apprPHII_1}
	\Phi_I (t) & = &
	\lambda(H) I_0 \Omega v_0^d
	\int_0^t m_0(t') (t - t')^d  \rmd t' \nonumber \\
  & \approx &
	\lambda(H) \frac{(d+1)\ln (z_0)}{\tau_0^{d+1}}
	|m_{\text{st},0}|
	\Biggl\{
	-  \int_0^t (t - t')^d  \rmd t'
	\nonumber \\ & & \label{eq:apprPHII_2}
	+ z_0 \int_0^t
		\exp \left[ - \ln (z_0) \left( \frac{t'}{\tau_0}
		\right)^{d+1} \right]
		(t - t')^d \rmd t' \Biggr\} \nonumber \\
& = & \lambda(H) |m_{\text{st},0}|
	\Biggl\{ - \Phi_0 (t) +
		\frac{z_0 (d+1)\ln (z_0)}{\tau_0^{d+1}}
	\nonumber \\ & & \label{eq:apprPHII_3}
	\times \int_0^t \sum_{k=0}^d
		\pmatrix{d \cr k} t^k
		\left( -t' \right)^{d-k}
		\exp \left[ - \ln (z_0) \left( \frac{t'}{\tau_0}
		\right)^{d+1} \right]  \rmd t'
	\Biggr\} \nonumber \\
& = & \lambda(H) |m_{\text{st},0}| \Bigglb(
		z_0 \left\{ \sum_{k=0}^d
		\pmatrix{d \cr k} (-1)^{d-k}
		\left[ \Phi_0 (t) \right]^{k/(d+1)}
		\int_0^{\Phi_0 (t)}
		x^{-k/(d+1)}
		\rme ^{-x} \rmd x \right\}
		\nonumber \\ & & \label{eq:apprPHII_4}
		- \Phi_0 (t) \Biggrb) \nonumber \\
& = & \lambda(H) |m_{\text{st},0}| \Bigglb(
		z_0 \left\{ \sum_{k=0}^d
		\pmatrix{d \cr k} (-1)^{d-k}
		\left[ \Phi_0 (t) \right]^{k/(d+1)}
		\gamma \left( 1 - \frac{k}{d+1}, \Phi_0 (t) \right)
		\right\}
		\nonumber \\ & & \label{eq:apprPHII_5}
		- \Phi_0 (t)
		\Biggrb) \; .
\end  {eqnarray}

Combining Eq.~(\ref{eq:apprPHIV_3}),
Eq.~(\ref{eq:apprPHII_5}), and Eq.~(\ref{eq:m1}) gives $m_1(t)$.

\bibliographystyle{prsty}

\newpage
\vspace*{5.2in}
\newpage
\vspace*{5.2in}
\newpage
\newpage
\newpage
\newpage

\end{document}